\documentclass[notitlepage,showkeys,showpacs]{revtex4-1}
\usepackage{graphicx}
\usepackage{amsmath}
\usepackage{amsfonts}
\usepackage{amssymb}
\usepackage{tensor}
\usepackage{enumerate}
\newcommand\scalemath[2]{\scalebox{#1}{\mbox{\ensuremath{\displaystyle #2}}}}

\newcommand{\be}{\begin{equation}}
\newcommand{\ee}{\end{equation}}
\newcommand{\bh}{\bar{h}}

\begin{document}
\title{Induced Perturbations and Stochastic effects in Collapsing Relativistic
 Stars} 
\date{\today}
\author{ Seema Satin} 
\affiliation{Indian Institute for Science Education and Research, Mohali, 
India }
\email{satin@iisermohali.ac.in }
\begin{abstract}
We present a modified model for relativistic stars which are usually 
represented by
perfect fluids. Fluctuations of the stress tensor act as source in the 
modified Einstein's equation, giving it a Langevin equation form. The
 occurence of these fluctuations is attributed to the microphysics of the
 interior of the star and their contribution  to statistical
 properties of the fluid and the induced metric perturbations are argued to be
 of significance.  We also discuss the response 
of  fluctuations of the stress tensor in the interior of the star and
 possible developments towards fluctuation-dissipation theorem issues in
 curved spacetime.  The aim  and further directions of research envisioned are
discussed intermittently throughout the manuscript and towards the end.   
\end{abstract}
\maketitle
\section{Introduction}
Perturbations in relativistic stars \cite{hartle,chandra1,kip,friedman} and
 black holes \cite{regge1,chandra2,clark, sengupta2} comprise a vast
 literature, addressing stability issues and oscillations as in  
 the area of asterosiesmology and singularities formed as end states of
 gravitational collapse of massive stars.

These compact configurations of relativistic stars are often modeled by perfect fluid as the matter constituent. 
In this article we attempt to raise a new aspect, that of taking into 
account, the fluctuations of the classical stress tensor and its
 effects in the interior of the star as well as, backreaction on the
 spacetime geometry in relativistic case.
Cowling approximation \cite{friedman, cowling} is often used in asteroseismology (both
the non-relativistic and relativistic cases) to ignore the effects of matter
perturbations on the spacetime geometry or metric. However this  is limited to
 few specific conditions and certain modes of oscillations and frequencies. 
  We give a full general relativistic
 treatment of the equations governing perturbations and their backreaction
 effect on the metric, including stochasticity and do not render the Cowling
approximation suitable for a general perturbative analysis as presented here . 
   This is so, because our interest is in exploring the effects of the 
 perturbations and stochastcity in a different context. 

In order to include stochasticity of the interior of a relativistic star
 and look at its backreaction consistently, the idea of a Langevin Equation
 formalism is adopted from  semiclassical stochastic
gravity \cite{bei1,bei2}. The semiclassical Einstein-Langevin equation takes
 account of the fluctuations of the
 quantum stress tensor and its backreaction, with interesting applications
 to early universe cosmology \cite{roura,bei3} and black hole physics
 \cite{sukanya, seema1,seema2}. 
We borrow the idea of setting up a Langevin approach in the classical
domain from the semiclassical case. However this has a very different 
 basic formalism  regarding mathematical and physical issues to tackle
with for us, as well as different domain of applications. While semiclassical
 Einstien Langevin equation 
is shown to be of importance to cosmology and quantum aspects of
black hole physics, the classical counterpart can be seen to cover the
domain of relativistic astrophysics, that of  stellar dynamics , 
asterioseismology and collapse of relativistic stars.  

A perfect fluid model, is often the simplest case considered for a
 relativisitic star. The microscopic particles in the perfect fluid collide
 frequently and their mean free path is short compared to the scale on
which density changes. Thus the stress tensor can be defined in  elements 
of the fluid which are small compared to macroscopic length scale but large 
compared to mean free path. Hence, it is the mean stress
tensor defined by $T^{ab}(x)$, with the 
pressure and density as averaged macroscopic quantities that is usually 
considered. 
The stress tensor is given by 
\be \label{eq:stress}
T^{ab} = u^a u^b (p+ \epsilon) + g^{ab} p
\ee 
 Here $p$ denotes pressure and $\epsilon$ the matter energy  density. 

 The average stress tensor thus descibed leaves scope to
define fluctuations given by $T^{ab}(x) - < T^{ab}(x)> = \tau^{ab}(x)$. 
 These may  be sourced via various processes in the interior of the star
 including quantum effects which are partially captured by 
fluctuations in the density and pressure, the effects being of
 classical stochastic nature. 
  This  model can be considered in terms of small deviation from that of a
 perfect fluid . 

There  can be various additional effects including imperfect and dissipative
 fluid models \cite{friedman},  for the study of  more realistic 
relativistic stars.  In these models it may be appropriate to associate the
 fluctuations with dissipative mechanism and thermal effects explicitly. 
 Our formalism here can be extended very easily to all these cases.
However in this article, we restrict ourselves to the stress tensor for
 a perfect fluid as the simpest example  in order to set up
the basic formalism correctly. 

\section{The Model using Einstein Langevin equation }
A  nonrotating spherically symmetric star in schwarzchild coordinates 
is given by  

\be
ds^2 = - e^{2 \nu(r,t)} dt^2 + e^{2\psi(r,t)} dr^2 + r^2 d \Omega
\ee

To consistently take into account the fluctuations of the stress tensor
 for a general relativistic treatment an Einstein Langevin equation  
can  be written as
\be \label{eq:pert}
G^{ab}[g+h](x)   =  T^{ab}[g+h;\xi](x) + \tau^{ab}[g](x) 
\ee

where $\tau^{ab}(x)$ on the background spacetime $g^{ab}(x)$ is defined
 stochastically such that $<\tau^{ab}(x)> =0$
and is covariantly conserved, $\nabla_a \tau^{ab}(x) =0 $. This follows 
from the covariance of $T^{ab}(x) $ . 
Perturbations  are defined to be deformations in the fluid elements, by way
 of shift from the equilibrium configuration and are deterministic in nature.
These perturbations in  spacetime and fluid variables can be described by 
 $h_{ab}$ and Lagrangian dispacement vector field $\xi^a $.

 One can  then see that the standard form of  Einstein's Equation on averaging
of the above equation is obtained.
Equation (\ref{eq:pert}) can be shown to be gauge 
invariant under the change of  metric perturbations given by $h_{ab}' = 
h_{ab} + \nabla_a \zeta_b + \nabla_b \zeta_a $, where
$\zeta^a$ is a stochastic  vector field on the background manifold. 
It reduces to the following perturbed form, 
\be
\delta G^{ab}[h](x) = \delta T^{ab}[h;\xi](x) + \tau^{ab}[g](x)
\ee
The trace reversed metric perturbation defined as
\[\bar{h}_{ab} = h_{ab} - \frac{1}{2} g_{ab} h  \mbox{ ; }  
h_{ab} = \bar{h}_{ab} -\frac{1}{2} g_{ab} \bar{h}
\]
 satifies 
\be
\nabla_a \bar{h}^{ab} = 0 
\ee
 We use this gauge to  get a  more desirable form of the Einstien Langevin
 equation which reads 
\be \label{eq:elangevin1} 
 -\frac{1}{2} \Box \bh^{ab}(x) + \tensor{A}{^{ab}_{cd}}(x) \bh^{cd}(x) =    
\tensor{W}{^{ab}_{cd}}(x) \bh^{cd}(x) + \tensor{E}{^{ab}_{cd}}(x)
 (\nabla^c \xi^d(x) + \nabla^d \xi^c(x)) 
+ \mathcal{L}_\xi T^{ab}(x) +  \tau^{ab}(x) 
\ee
where
\be
 \tensor{W}{^{ab}_{cd}}(x) = \tensor{E}{^{ab}_{cd}}(x)- \frac{1}{2}
 \tensor{E}{^{abef}}(x)g_{ef}(x) g^{cd}(x)
\ee
 The lhs of  equation (\ref{eq:elangevin1}) is written using the 
Einstein tensor perturbations, while the rhs is that of the stress tensor.
The tensors $\tensor{A}{^{ab}_{cd}} $  and $\tensor{E}{^{ab}_{cd}}$ are
 given in the appendix, while
\[ \mathcal{L}_\xi T^{ab}(x) = \nabla_c (T^{ab} \xi^c) - T^{cb} \nabla_c \xi^a
- T^{ac} \nabla_c \xi^b \]
One can decipher the tensor $E^{abcd}(x)$   with dissipation 
in the fluid . This can be shown as arising due to response of the fluctuations
 causing induced perturbations in the fluid. This is discussed  in more detail
 later.  An important point
to notice here is that, dissipation seems to be local, unlike many cases
where usually a nonlocal dissipation comes out naturally, (including as
 obtained in semiclassical stochastic gravity ). Also
it is unlike diffusion constant for ordinary brownian motion theory in
 Newtonian case. Thus one may see an example of a local dissipative
 effect here, which can be interesting  for statistical  properties  of the
 system.

Now Einstein Langevin equation takes the form,
\be \label{eq:elangevin} 
 -\frac{1}{2} \Box \bh^{ab} = [-\tensor{A}{^{ab}_{cd}} +  
  \tensor{W}{^{ab}_{cd}}]\bh^{cd} + D^{ab}(x) + \tau^{ab} 
\ee
where we have put,
\be
 D^{ab}(x) = E^{abcd}(x) (\nabla_c \xi_d(x) +
 \nabla_d \xi_c(x) ) - \mathcal{L}_\xi T^{ab}(x) 
\ee
 The aim of such a formalism is to lay a  background for
 doing non equilibrium or near equlibrium statistical physics of 
  isolated relativistic stars during different stages of collapse or 
including near equilibrium states.

 As can be seen, $D^{ab}$ arises due to $\xi^a$, we can associate
 the perturbative effect in the fluid as a response to the
stochastic fluctuations.  
\section{ Form of solution of the Einstein Langevin Equation} 
 Proceeding towards the issues of backreaction of the Langevin noise,
equation (\ref{eq:elangevin})  can be written in the following form  
\be
[\frac{1}{2} \delta^a_c(x) \delta^b_d(x) \Box + \tensor{A}{^{ab}_{cd}}(x) - 
\tensor{W}{^{ab}_{cd}}(x) ] \bh^{cd}(x) = D^{ab}(x) + \tau^{ab}(x)
\ee
A general form of  solution is thus given by,
\be \label{eq:13} 
\bh^{cd}(x) = \int G^{abcd}(x,x') [ D_{ab}(x') + \tau_{ab}(x') ] dx'
\ee 
where $\tensor{G}{^{ab}_{cd}}(x,x')$ is the Green's function for the operator
$[\frac{1}{2} \delta^a_c \delta^b_d \Box + \tensor{A}{^{ab}_{cd}} - 
\tensor{W}{^{ab}_{cd}}]$.

 As is known that Green's function in curved spacetime is not easy to
get in analytical form and only few very specific examples have been obtained
e.g \cite{cadwell} for specialised cases. Since here 
our intention is to demonstrate the contribution of stochasticity to the 
perturbations, 
we do not go forward to develop methods for obtaining the Green's function in 
explicit form and leave it as a separate exercise for further developments,
 which is quite an invovled task in itself.
 We may also later looking at the details of the coupled differential 
equations that we get here, decide if numerical methods are 
better suited for a complete solution. This is yet to be investigated and we
 make no claims here as to the general method of solution for the 
Einstein Langevin equations, since these many vary from case to case. Rather
we give a general form of solution which brings out clearly the
two point correlations and stochastic nature of the same. 

From equation (\ref{eq:13}) the two point correlation of the metric perturbations can be written as   
\be \label{eq:twopoint}
<\bh^{ab}(x) \bh^{cd}(y)> = \int \int  G^{ablm}(x,x') G^{cdpq}(y,y')
 [ D_{lm}(x') D_{pq}(y') +  N_{lmpq}(x',y') ] dx' dy'
\ee
where
\be
 N_{lmpq}(x',y') = < \tau_{lm}(x') \tau_{pq} (y') >   
\ee
is the noise kernel or two point correlation of the stress tensor fluctuations.
Towards the end , we discuss more about noise in the system.

From equation (\ref{eq:13}) one can also see that,  
\be
<\bh^{cd}(x)> = \int G^{abcd}(x,x') D_{ab}(x') dx'
\ee
 which clearly shows the overall contribution of the fluctuations on metric
 perturbations seems to be vanishing. While this looks 
 to be quite a consistent result physically, the importance of including the
 fluctuations in the Einstein's Equations  can be
 seen clearly in the two point (or higher) correlations 
(\ref{eq:twopoint}). 

In the coincidence limit equation (\ref{eq:twopoint}) takes the form
\be 
<\bh^{ab}(x) \bh^{cd}(x)> = \int \int  G^{ablm}(x,x') G^{cdpq}(x,y')
 [ D_{lm}(x') D_{pq}(y') +  N_{lmpq}(x',y') ] dx' dy'
\ee
 Hence though the explicit contribution of the stochastic term seems to be
 vanishing on the mean $ <h^{ab}(x)> $, it is not so for the rms of $h^{ab}$.

Here we attempt to set up the basic theoretical structure, 
  it will be future endeavour to relate these quantities
 to observations in  asterioseismology and oscillations of relativistic stars
and have estimates for the magnitude of the contributions. 

Now we show more explicitly, the perturbed stochastic equations in comoving
coordinates and the form of solutions. Comoving coordinates are 
 useful in modeling collapsing stars, and hence we are interested in 
getting the expressions in these coordinates. 
\subsection{Expressions in Comoving Coordinates}
In comoving coordinates a spherically symmetric (non-rotating) relativistic
star is given by
\be
ds^2 = - e^{2 \nu(r,t)} dt^2 + e^{2\psi(r,t)} dr^2 + R^2(t,r) d \Omega
\ee
where $R(t,r)$ is the area radius and $t$ and $r$ are the comoving time
and radius.
The energy momentum tensor for a perfect fluid in these coordinates is of
 the form:
\be \label{eq:stress}
T^{ab}(x) = \left(
\scalemath{0.6}{
\begin{array}{cccc}
 g^{00}(x) \epsilon(t,r) & 0 & 0 & 0 \\
0 &  g^{11}(x) p(t,r) & 0 & 0 \\
0 & 0 &  g^{22}(x) p(t,r) & 0 \\
0 & 0 & 0 & g^{33}(x) p (t,r)
\end{array}
 } 
\right)
\ee
 We assume $p_r = p_\theta = p_\phi $ for a simple model, and the velocity
vector of the comoving observer is given  by $u^a = (1,0,0,0)$ which satisfies
the condition $u^au_a = -1 $ .

Here  the Lagrangian change for a quantity is denoted by $\Delta z$ and
 Eulerian change by $\delta z $ such that,
\be
\delta z = \Delta z - \mathcal{L}_\xi z = \Delta z - \xi^\alpha \nabla_\alpha z
\ee
Thus $ \Delta g_{ab} = h_{ab} + \nabla_a \xi_b + \nabla_b \xi_a $.
The perturbation in pressure and density are given by 
\begin{eqnarray}
\delta \epsilon & = & - \frac{1}{2} (\epsilon+ p) q_{cd} \Delta g^{cd} -
 \xi.\nabla \epsilon \nonumber  \\
\delta p & = & -\frac{1}{2} \Gamma_1 p q_{cd} \Delta g^{cd} - \xi.\nabla p
\end{eqnarray}
where $q_{ab} = u_a u_b + g_{ab}$ and $\Gamma_1$ is the usual  adiabatic index.
We consider only radial perturbations here and hence only $\xi^r$ component
 for the Lagrangian displacement vector being non-zero, while taking only
spherically symmetric metric perturbations given by, 
 $h^{00}(x), h^{11}(x), h^{22}(x) $ and $h^{33}(x)$ to be the
non-zero for  simplicity of solution. 
The Einstein Langevin equation then gives the following set,
\begin{eqnarray} \label{eq:solel}
\tilde{F}_1 \bh^{00} & = & a_1 \bh^{11} + a_2 \bh^{22} + a_3 \bh^{33} -
g^{00} f_1 + \tau^{00} \nonumber \\
\tilde{F}_2 \bh^{11} & = & b_1 \bh^{00} + b_2 \bh^{22} + b_3 \bh^{33} -
g^{11} f_2 + \tau^{11} \nonumber \\
\tilde{F}_3 \bh^{22} & = & c_1 \bh^{00} + c_2 \bh^{11} + c_3 \bh^{33} 
- g^{22} f_2 + \tau^{22} \nonumber \\
\tilde{F}_4 \bh^{33} & = & d_1 \bh^{00} + d_2 \bh^{11} + d_3 \bh^{22} -
g^{33} f_2 + \tau^{33}
\end{eqnarray}  
All the coefficients and operators in the equation are given in the appendix
 and
\[
 f_1  =    (\epsilon + p) \xi^{r'} + \epsilon ' \xi^r  ;
f_2  =   p \Gamma_1 \xi^{r'} + \xi^r p'  
\]
The above set of coupled  equations can be solved numerically in general,
but here we would be interested in two point correlations of the metric and
solution in terms of Langevin  approach. 
In an upcoming article our effort will be to get a complete formal 
stochastic solution of the above using the Regge Wheeler gauge. 

 We further assume $\bh^{22} = \bh^{33}$.  
Substituting the expression for $\bh^{22}$ obtained from the last two equations
of (\ref{eq:solel}), in the first two equations of the set, we obtain 
\begin{eqnarray}
\tilde{D}_1 \bh^{00} (x) & = & k_1(x) \bh^{11}(x) + m_1(x) + k_2(x) 
( \tau^{22}(x) - \tau^{33}(x) ) + \tau^{00}(x) \nonumber \\
\tilde{D}_2 \bh^{11} (x)  & = & p_1(x) \bh^{00}(x) + m_2(x) + p_2(x) 
(\tau^{22}(x) - \tau^{33}(x)) + \tau^{11}(x) 
\end{eqnarray}
The operators and coefficients can be verified from the appendix.
We show below the expression for $\bh^{00}$, a similar expression for 
$\bh^{11}$ can be obtained easily. Thus
\be
[\tilde{D}_2 ( \frac{1}{k_1} \tilde{D}_1) - p_1 ] \bh^{00} = Q_1(x) 
 + Q_2(x)( \tau^{22} - \tau^{33}) +  Q_3(x) \tau^{00} + \tau^{11}
\ee   
where
\[ Q_1 (x) = - \tilde{D}_2 (m_2/k_1) + m_2 ; Q_2(x) = -\tilde{D}_2(k_2/k_1)
 + p_2 ; Q_3(x) = - \tilde{D}_2 (1/k_1) \]
 
The final form of $\bh^{00}(x) $ is then obtained as, 
\be
 \bh^{00}(x) = \int G_1(x,x')[ Q_1(x') + Q_2(x')( \tau^{22}(x') - \tau^{33}(x')
) +  Q_3(x') \tau^{00}(x') + \tau^{11}(x')] d^4 x'
\ee   
where $G_1(x,x')$ is the Green's function for the operator $\{ \tilde{D}_2 
(\frac{1}{k_1} \tilde{D}_1) - p_1 \} $.
The mean of the  metric perturbation is then given by
\be \label{eq:av}
<\bh^{00}(x)> = \int G_1(x,x') Q_1(x') d^4x'
\ee
we see a vanishing contribution of the fluctuation terms here as expected. There
is an overall smearing effect which is physically desirable for such a
stochasticity. Though the stochastic term itself vanishes, but we see that
 it induces an effect (in equation (\ref{eq:av})) given by $Q_1(x')$ 
 in the fluid variables  comprising of the Lagrangian dispacement $\xi^r$
 and the metric perturbations in the gravity sector. 
 Here we emphasise that there is no external agency giving rise to
these perturbative effects in the relativistic star. In general theory
of perturbations for stars, there is either an internal rotational 
effect of the fluid or and external agency through which the perturbations
and thus oscillations are induced. Here the physical picture
is drastically different, in that, the stochastic effects inside the
 star give rise to similar perturbative effects as can be seen to arise 
due to the former  mentioned sources. This is consistent with the linear 
response theory in statistical physics. Hence we propose that even in 
isolated relativistic stars, either undergoing dynamical collapse or
  near equilibrium configurations of compact configurations, one
can  see these perturbative effects, which can give rise to oscillations 
inside the star. 
 We do not  at present claim any estimate of the frequency of oscillations or
magnitude of pertubations, but  assume these to be strong enough to
have backreaction effects. The significance of even the slightest
disturbances or low magnitude of these perturbations, both in the
fluid and the spacetime geomtry can have  consequences near critical
phases of collapse,where they may affect  important,
 decisive physical results of the collapsing cloud. 
Further, the correlations of the metric perturbations thus obtained ,  
near critical points are expected to enable one to  do  statistical analysis
 of spacetime geometry near these phases.
 This is one of our the prime aims, (along with study of oscillations
set by the stochastic effects) in establishing such a formalism as in this
 article.    

One can see that, the two point correlation takes the form 
\be
<\bh^{00}(x) \bh^{00}(y) > = \int \int G_1(x,x') G_1(y,y') [ Q_1(x')
Q_1(y') + S(x',y') ] dx' dy'  
\ee
here $S(x',y')$ is the stochastic part and is given in the appendix.
While in the coincident limit we  would get the mean square of the
 metric perturbations and can be easily followed from the above expression
by taking $x \rightarrow y$. 
\section{Noise, Fluctation and Dissipation}
 Below we discuss a few aspects of noise.

The two point correlation of the fluctuations of the stress
 tensor can be given by 
\be
N^{abcd}(x,y) = < \tau^{ab}(x) \tau^{cd}(y) > = N^{cdab}(y,x)
\ee
from the definition of $\tau^{ab}(x)>$ it follows that, 
\be
N^{abcd}(x,y) = < T^{ab}(x)T^{cd}(y) > - <T^{ab}(x)>< T^{cd}(y)>
\ee
Since $\nabla_a \tau^{ab}=0 $, as mentioned earlier, the noise satistifes the
 following property,
\be
\nabla_a N^{abcd}(x,y) = \nabla_b N^{abcd}(x,y) = \nabla_c N^{abcd}(x,y)
 = \nabla_d N^{abcd}(x,y) =0
\ee 
For the spherically symmetric spacetime and stress tensor in comoving
 coordinates given by equation (\ref{eq:stress}), the  non-zero components of 
the noise  are  can be obtained as below.
\begin{eqnarray}
N^{0000}(x,x') &=& e^{-2(\nu(t,r)+\nu(t',r'))}[<\epsilon(t,r) \epsilon(t',r')> -
 <\epsilon(t,r)> <\epsilon(t',r')>] \nonumber \\
N^{0011}(x,x') &=& -e^{-2(\nu(t,r)+ \psi(t',r'))}[<\epsilon(t,r) p(t',r')> -
 <\epsilon(t,r)> <p(t',r')>]
 \nonumber \\
N^{1100}(x,x') &=&- e^{-2(\psi(t,r)+\nu(t',r'))}[<p(t,r) \epsilon(t',r')> -
 <p(t,r)> <\epsilon(t',r')>] \nonumber \\
N^{0022}(x,x') & = & \frac{-e^{-2\nu(t,r)}}{R^2(t',r')}[<\epsilon(t,r)
 p(t',r')> - <\epsilon(t,r)> <p(t',r')>] \nonumber \\
N^{2200}(x,x') &=& \frac{-e^{-2 \nu(t',r')}}{R^2(t,r)}[<p(t,r) 
\epsilon(t',r')> - <p(t,r)> <\epsilon(t',r')>] \nonumber \\
N^{0033}(x,x') & = & \frac{- e^{-2\nu(t,r)}}{R^2(t',r')\sin^2\theta'}[<\epsilon
(t,r) p(t',r')> - <\epsilon(t,r)> <p(t',r')>] \nonumber \\
N^{3300}(x,x') & = &\frac{-e^{-2\nu(t',r')}}{R^2(t,r)\sin^2\theta)} [<p(t,r) 
\epsilon(t',r')> - <p(t,r)> <\epsilon(t',r')>] \nonumber \\
N^{1122}(x,x') &=& \frac{e^{-2 \psi(t,r)}}{R^2(t',r')}[<p(t,r) p(t',r')> -
 <p(t,r)> <p(t',r')>] \nonumber \\
N^{2211}(x,x')&=& \frac{e^{-2\psi(t',r')}}{R^2(t,r)}[<p(t,r) p(t',r')> -
 <p(t,r)> <p(t',r')>] \nonumber \\
N^{1133}(x,x') &=& \frac{e^{-2 \psi(t,r)}}{R^2(t',r') \sin^2 \theta'}[<p(t,r)
 p(t',r')> - <p(t,r)> <p(t',r')>] \nonumber \\
N^{3311}(x,x') &=& \frac{e^{-2\psi(t',r')}}{R^2(t,r)\sin^2\theta)}[<p(t,r)
 p(t',r')> - <p(t,r)> <p(t',r')>] \nonumber \\
N^{2233}(x,x') &=& \frac{1}{R^2(t,r) R^2(t',r') \sin^2 \theta'}[<p(t,r) 
p(t',r')> - <p(t,r)> <p(t',r')>] \nonumber \\
N^{3322}(x,x') &=& \frac{1}{R^2(t,r) \sin^2 \theta R^2(t',r')}[<p(t,r)
 p(t',r')> - <p(t,r)> <p(t',r')>] \nonumber \\
N^{1111}(x,x')&=& e^{-2(\psi(t,r)+ \psi(t',r'))}[<p(t,r) p(t',r')> - <p(t,r)>
 <p(t',r')>] \nonumber \\
N^{2222}(x,x') &=& \frac{1}{R^2(t,r) R^2(t',r')}[<p(t,r) p(t',r')> - <p(t,r)>
 <p(t',r')>] \nonumber \\
N^{3333}(x,x') &=& \frac{1}{R^2(t,r) R^2(t',r') \sin^2 \theta \sin^2 \theta'}
 [<p(t,r) p(t',r')> - <p(t,r)> <p(t',r')>] 
\end{eqnarray} 

The dissipation kernel for the same geometry and stress tensor
can be shown to have the following components.
\begin{eqnarray}
& & E^{0000}(x) = \frac{1}{2}\{ \epsilon(t,r) + p(t,r)(1+ e^{-4 \nu(t,r)}) \}
 \mbox{ , }
E^{0011}(x) = \frac{1}{2} p(t,r) e^{-2(\nu(t,r) + \psi(t,r))} \mbox{ , } 
E^{0022}(x) = \frac{1}{2} p(t,r) \frac{e^{-2 \nu(t,r)}}{R^2(t,r)}\nonumber \\
& & E^{0033}(x) = \frac{1}{2} p(t,r) \frac{e^{-2 \nu(t,r)}}{R^2(t,r) \sin^2 
\theta} \mbox{ , } 
E^{1111}(x) = - \frac{1}{2} p(t,r) e^{-4 \psi(t,r)} (1 + \Gamma_1)
\mbox{ , } E^{1122}(x) = \frac{1}{2} \frac{e^{-2 \psi(t,r)}}{R^2(t,r)} p(t,r
) (1-\Gamma_1) \nonumber \\
& &  E^{1133}(x) = \frac{1}{2} p(t,r) \frac{e^{-2\psi(t,r)}}{R^2(t,r)
 \sin^2 \theta} (1- \Gamma_1) \mbox{ , } 
 E^{2222}(x) = - \frac{1}{2}\frac{p(t,r)}{R^4(t,r)} (1 + \Gamma_1) \mbox{ , }
E^{2233}(x) = \frac{1}{2} \frac{p(t,r)}{R^4(t,r) \sin^2 \theta} (1- \Gamma_1)
\nonumber \\
& &  E^{3333}(x) = - \frac{1}{2} \frac{p(t,r)}{ R^4(t,r) \sin^4 \theta}
 (1 + \Gamma_1)
\end{eqnarray}
The nature of dissipation being local is certainly interesting, and we
would attempt to find out consequences of the same in the statistical properties
of a collapsing relativistic star in future work.

A fluctuation dissipation theorem can be written as
\be
N^{abcd}(x,x') = \int K(x,x'; y) E^{abcd}(y) dy
\ee
here $K(x,x';y)$ is the fluctuation-dissipation kernel. The locality of the
disspation kernel here, may thus lay some restriction on the model of noise
that is physically allowed for such systems. This needs further investigations.
\section{Conclusions}
As concluding remarks, we highlight few additional issues that can be raised
over the stochasticity introduced here in the model for relativistic stars.

 A collapsing star goes through different
 phases, where classical and quantum effects become important
to study the interior regions of the star and overall dynamics.
 The formulation presented in this manuscript may also be useful for
investigations in the following directions. 

  One can try to explore if there is  mesoscopic regime between
 classical and quantum  effects in the intermediate phases of
 gravitational collapse. 
These  classical fluctuations, then can  take over before
 quantum effects set in as the singularity is reached, and can lead to
short term where mesoscopic physics becomes important during the collapse of
 the star. 
In additon to this, stochasticity  in the strong field regime  due to 
 microscopic or partially captured quantum effects can be probably linked  to
 graviational decoherence \cite{bassi} which is an 
upcoming area.  This would be relevant towards the end states of collapse
 which result either in singularities  or more exotic compact objects rather
 than stable neutron stars. 
\begin{acknowledgements}
The author is thankful to T.Padmanabhan, Bei Lok Hu and Jasjeet Bagla for
 helpful discussions and to IUCAA Pune, India, for offering a visiting position
where part of the work was planned.
\end{acknowledgements}
\appendix
\section*{Appendix}
\begin{eqnarray}
\tensor{A}{^{abcd}}& = &[ \frac{1}{2} \{ g^{ab} R^{cd} + g^{cd} R^{ab} -
 g^{ac} R^{bd} - g^{ad} R^{bc} + ( g^{ac} g^{bd} - \frac{1}{2} g^{ab} g^{cd})\}
- 2 R^{acbd} ]  \nonumber \\
\tensor{E}{^{abcd}}&  =& \frac{1}{2} (\epsilon + p ) u^a u^b u^c u^d +
\frac{1}{2} p ( g^{ab} g^{cd} - g^{ac} g^{bd} - g^{ad} g^{bc} ) 
 - \frac{1}{2} \Gamma_1 p q^{ab} q^{cd}  \nonumber  \\
\tilde{F}_1 & = & \{-\frac{1}{2} \Box + \frac{1}{4}(\epsilon+ 3p) +
 \tensor{A}{^{00}_{00}} \}
\nonumber \\
\tilde{F}_2 & = & \{-\frac{1}{2} \Box - \frac{1}{2}(1-\frac{1}{2} \Gamma_1)p +
 \tensor{A}{^{11}_{11}} \} \nonumber \\
\tilde{F}_3 & = & \{-\frac{1}{2} \Box +\frac{1}{2}( 1- \frac{1}{2}\Gamma_1)p +
 \tensor{A}{^{22}_{22}} \} \nonumber \\
\tilde{F}_4 & = & \{-\frac{1}{2} \Box +\frac{1}{2}(1-\frac{1}{2} \Gamma_1) p+ 
 \tensor{A}{^{33}_{33}} \} \nonumber \\
 & & a_1  = - \tensor{A}{^{00}_{11}} - g^{00} g_{11}\frac{1}{4}(3\epsilon+p)
 \mbox{ , } a_2 =
 - \tensor{A}{^{00}_{22}} - g^{00} g_{22}\frac{1}{4}(3\epsilon+ p)  \nonumber\\ 
  & & a_3 = -\tensor{A}{^{00}_{33}} - g^{00} g_{33}\frac{1}{4}(3\epsilon+p)  
 \mbox { , } b_1 =- \tensor{A}{^{11}_{00}}- \frac{1}{2} g_{00} g^{11} p\Gamma_1
 p(1+\frac{3}{2} \Gamma_1)   \nonumber \\
& &  b_2 = -\tensor{A}{^{11}_{22}}  + \frac{1}{2} g^{11} g_{22} p
 (1+ \frac{\Gamma_1 }{2}) \mbox{ , }
b_3 = -\tensor{A}{^{11}_{33}}+ \frac{1}{2} p g^{11} g_{33}
 (1+ \frac{1}{2} \Gamma_1).  \nonumber \\
& & c_1 = -\tensor{A}{^{22}_{00}}+ g^{22} \frac{1}{2} (1+\frac{3}{2}\Gamma_1) g_{00} \mbox{ , }
 c_2 = -\tensor{A}{^{22}_{11}}+ g^{22} \frac{1}{2}(1+ \frac{\Gamma_1}{2}) p
 g_{11}   \nonumber \\ 
 & & c_3 = -\tensor{A}{^{22}_{33}}- g^{22} \frac{1}{2} (1+\frac{\Gamma_1}{2}) p
 g_{33}   \mbox{ , } 
d_1 = - \tensor{A}{^{33}_{00}}+g^{33} \frac{1}{2} (1+\frac{3}{2}\Gamma_1) p
 g_{00}  \nonumber \\ 
  & & d_2 = -\tensor{A}{^{33}_{11}}+ g^{33} \frac{1}{2}(1+\frac{\Gamma_1}{2}) p
 g_{11} \mbox{ , }
  d_3 = -\tensor{A}{^{33}_{22}}+ g^{33}\frac{1}{2}(1+\frac{\Gamma_1}{2}) p
 g_{22} \nonumber \\
 & & s_1  =  \tensor{A}{^{22}_{22}} + \tensor{A}{^{22}_{33}} + \frac{1}{2} p(1-
\frac{\Gamma_1}{2}) - \frac{1}{2} p g^{22} g_{33} \mbox{ , } 
s_2  =  - \tensor{A}{^{22}_{00}} + \frac{1}{2} g^{22} g_{00} p(1+ \frac{3}{2}
\Gamma_1)  \nonumber \\ 
 & & s_3  =  - \tensor{A}{^{22}_{11}} + \frac{1}{2} g^{22} g_{11} ( 1 + 
\frac{\Gamma_1}{2})  \mbox{ , } 
 l_1  =  p + \tensor{A}{^{33}_{22}} + \tensor{A}{^{33}_{33}} - \frac{1}{2} 
(1+ \frac{\Gamma_1}{2}) p  g^{33} (g_{22} + g_{33}) \nonumber \\
& & l_2   =   - \tensor{A}{^{33}_{00}} - \frac{1}{2} g^{33} p e^{2 \nu}
(1+ \frac{3}{2} \Gamma_1) \mbox{ , } 
l_3  =   -\tensor{A}{^{33}_{11}} + \frac{1}{2} g^{33} g_{11} p (1 +
\frac{\Gamma_1}{2})  \nonumber  \\ 
 & & k_1  =  a_1 + (a_2 + a_3) \frac{(s_2 - l_2)}{(s_1 - l_1)} \mbox{ , } 
k_2  =  \frac{(a_2 + a_3)}{(s_1 - l_1)} \mbox{ , } 
p_1  =  b_1 + (b_2 + b_3) \frac{(s_2 - l_2)}{(s_1 - l_1)}  \nonumber  \\
 & & p_2  =  \frac{(b_2 + b_3)}{(s_1 - l_1)}  \mbox{ , } 
\tilde{D}_1  =  \tilde{F}_1 - (a_2 + a_3) \frac{(s_2 - l_2)}{(s_1 - l_1)} 
 \mbox{ , }
\tilde{D}_2  =  \tilde{F}_2 - (b_2 + b_3) \frac{(s_3 - l_3)}{(s_1 - l_1)}  
  \nonumber   \\
 & & m_1  =  -g^{00} f_1 - k_2 f_2  \mbox{ , }
m_2  =  - (g^{11}+p_2) f_2  \nonumber \\ 
 S(x,y)  &  = &  Q_2(x) Q_(y) < \tau^{22}(x) \tau^{22}(y)> + < \tau^{33}(x)
\tau^{33}(y) > - < \tau^{22}(x) \tau^{33}(y) > - <\tau^{33}(x) \tau^{22}(y)> ]
  \nonumber  \\
  & & + Q_3(x) Q_3(y[ < \tau^{00}(x) \tau^{00}(y) > + Q_2(x) Q_3(y)[
 < \tau^{22}(x)\tau^{00}(y)> + <\tau^{22}(x) \tau^{11}(y)>  \nonumber \\
  & &  - <\tau^{33}(x) \tau^{00}(y)> 
 -< \tau^{33}(x) \tau^{11}(y)>] + Q_2(x) [ <\tau^{22}(x) \tau^{11}(y)> -
 <\tau^{33}(x) \tau^{11}(y)> ] \nonumber  \\
 & &   + Q_3(x) <\tau^{00}(x) \tau^{11}(y) > 
 + Q_2(y')[ <\tau^{11}(x) \tau^{22}(y) > ] - <\tau^{11}(x) \tau^{33}(y)>]
    \nonumber  \\
  & &  + Q_3(y) <\tau^{11}(x) \tau^{00}(y) > + < \tau^{11}(x) \tau^{11}(y) >
\nonumber 
\end{eqnarray}

\end{document}